## How to Deal with Fake News-Visualizing Disinformation

### Introduction

Aristotle reputedly said that humans by nature desire to know, and who could reasonably claim that their purpose in life is to know less? one of the defining features of a functional democracy is the discerning ability of its citizens to properly evaluate information that becomes part of the public discourse. In contemporary terms, that type of information is somewhat restricted due to the extensive specialization in all fields of knowledge, where the public's access is very limited in scope. Nevertheless, despite its shortcomings in that regard there is an assumption that such information has been properly disseminated in terms of the integrity of both its contents defined as its accuracy and precision, and the means of its propagation.

The importance of such considerations about the veracity and dissemination effectiveness of information lies squarely on the means of persuasion and manipulation that have become so readily available through mass media. In developing the means to examine how to discern between what is true and false in information, it is imperative to establish a distinction between misinformation and disinformation. Such a distinction can be made in terms of whether there is *agency,* and therefore *intent* present in the transmission of information. Misinformation is taken to lack such intent, whereas disinformation implicitly entails it. Misinformation can be present in a variety of settings from academic ones where the role of misconceptions can be detrimental to effective learning, to that of the public sphere where the context is progressive, and disinformation can spread much more easily. Studies have demonstrated that disinformation can spread much faster than the truth. The speed of propagation of disinformation compared to accurate information in the case of false news as opposed to true news approaches an order of magnitude, or roughly 10 times faster (Vosoughi et al, 2018).

There are several important issues to consider where disinformation propagation looms large. Among them are two that directly impact the effect on the general public's ability to properly deal with information, and they are both present in the context of conspiracies.

1) From a psychological perspective the effect of conspiracy theories on the public exhibits a peculiarly problematic feature. It has been found to be contrary to the default position, which should be one of skepticism (Clarke, 2002).
2) From a sociological perspective, it has been found that more than half of the US population believes in at least one conspiracy, and belief in one such theory quickly leads to espousal of other conspiracy theories, something that has significantly negative societal consequences (S. der Linden, 2015).

The objective of this essay is to propose a quantitative mechanism by which the spread of disinformation can be analyzed by modelling it as displaying its main features in terms of wave properties. Why is this significant? The use of concepts derived from wave phenomena can help us to understand properties of nature from the infinitesimally small, to the largest scales in the observable universe. In addition, there are many occurrences and events in a wide variety of experiences where information can be categorized as cycles or recurring instances of properties that can be understood in terms of those of waves. An understanding of wave motion can help us to describe phenomena apparently having nothing in common, in a way that enhances and promotes general knowledge. This is clearly the case in our attempts to deal with the spread of disinformation. The use of waves in the terminology used by the general population is quite common; however, going beyond its mere utility to describing and interpreting information as wave phenomena can be quite instructive. There are examples of applications of wave properties and behavior to areas beyond scientific contexts (Espinoza, 2017).

**Using Wave Properties**

We begin by introducing fundamental wave characteristics as these can be attributed to the propagation of disinformation. The progression will then be towards a phenomenological behavior in terms of its interaction with the public.

*Fundamental Characteristics*

Why are we compelled to use a wave as a representation? If disinformation propagates in a different way, such as material objects do, then several properties necessary for the interactive aspects introduced later will be absent. For example, mathematical modeling of information processing (Stewart et al., 2019) is impossible without including properties such as induction, diffusion, and distortion, processes analogous to refraction, absorption, superposition, and interference. Such processes are not associated with material objects, but they decidedly are with waves. In addition, the basic condition that gives rise to a wave representation can be applied to the spreading of disinformation since it is well documented that the number of times false news are repeated increases the likelihood of their credibility.

The very first feature deals with propagation, and that of course involves a description of disinformation as a wave, where the basic expression is that of its speed. The general expression for speed as the ratio of distance and time is given by v = $\frac{D}{t}$ where v is speed, D is distance, and t is time. The corresponding expression for a wave's speed is v = $\frac{\lambda}{T}$ = $\lambda f$ where $\lambda$ is the wavelength (the length of the wave, which corresponds to the distance), $T$ is the period (the time for one complete wave), and $f$ is the frequency (how often the wave pattern repeats).
We next designate each of these variables to parameters contained in disinformation propagation as a wave, along with Amplitude, and energy content.

Our next task is to obtain a representation of these characteristics as being those of a mechanical wave (one that requires material oscillations or vibrations as the components), and that can be either longitudinal (oscillations along the direction of propagation), or transverse (oscillations perpendicular to the direction of propagation).

How can each of these variables be represented?

1. The Amplitude of the wave represents the amount of disinformation being propagated.
2. The energy content is directly proportional to the intensity (Amplitude)$^2$ of the wave. The use of the square of the Amplitude is based on the variation of the amount of disinformation that occurs in both directions from an equilibrium or reference position (e.g., zero disinformation).

3. The wavelength represents the reach or scope along a chosen direction of propagation, where the varying amount of disinformation (the Amplitude) occurs in cycles.

4. The period represents the time for complete cycles where the Amplitude values repeat (e.g., maximum to maximum).

5. The frequency represents the number of times (how often) the disinformation contained in the wave repeats.

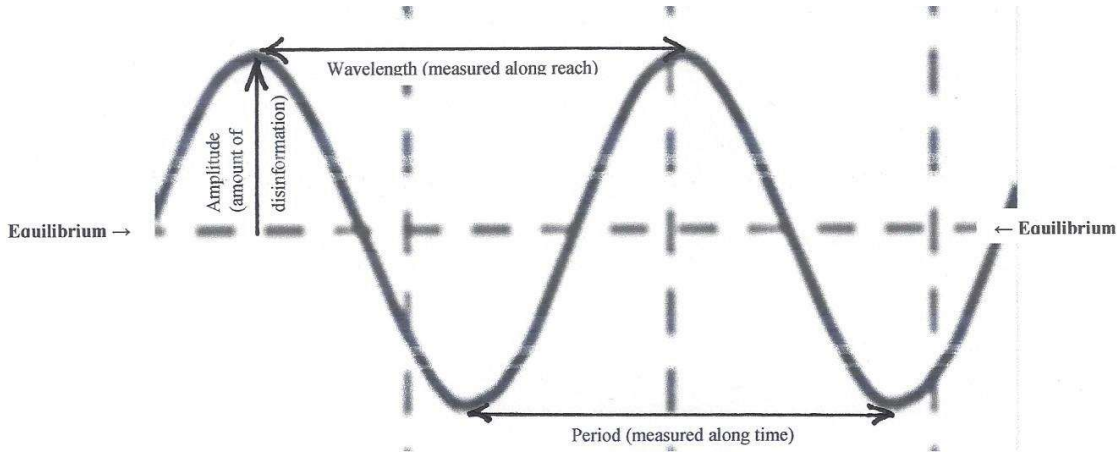

Figure 1. Wave characteristics as applied to disinformation.

How does the representation help us visualize the features of disinformation as an independently occurring phenomenon? Firstly, it should be obvious that if the Amplitude is zero (along the equilibrium line), there is no disinformation being propagated. As time goes on, the amount of disinformation begins to increase, reaching a maximum (the Amplitude), and then gradually decreasing (due to its unsustainability in terms of attention given to it). It follows a cycle where it recurs after a certain period, and it has a certain reach in terms of the number of believers. Incidentally, the Amplitude can also be expressed in terms of the number of believers, as waves can also have the same units vertically and horizontally. The frequency with which it occurs can be determined by visualizing that if the time for a cycle to repeat is long, the frequency is low. In other words, if the amount of disinformation is not repeated often, it will take a long time to reoccur; conversely if it has a high frequency of occurrence, it must have been repeated many times and/or in a shorter amount of time.

The speed of propagation can be expressed in terms of both the amount of reach (the wavelength), and the frequency (the number of times the disinformation is repeated). The dependence on frequency is particularly significant given the documented 'illusory truth effect' that helps explain why falsehoods propagate faster than the truth. The frequency of repetition of statements increases the perception of their truth, even if the statements are false (Fazio et al., 2015). Therefore, true statements are less likely to be repeated and there is a greater incentive in repeating falsehoods, thus resulting in a larger frequency and consequently a larger speed of propagation for disinformation. It is worth pointing out that in learning settings mere repetition of information can improve its retention and memorization, but it does not lead to meaningful comprehension. In the case of disinformation propagation, repetition is sufficient since the intent is to persuade without allowing for reflection, a necessary component of comprehension and understanding.

*Behavior and Interactivity*

Wave propagation leads to situations where several properties are subject to change. We consider the medium or material through which the waves propagate to be non-dispersive, meaning that there is only one value of the speed. However, as waves travel through a medium, and across different media some of the wave properties will change. For our purposes in representing the propagation and interaction of disinformation as waves interacting with objects and with each other, we shall concentrate on changes in speed and intensity. The objective is to relate the changes in these properties or variables, to the dissemination and impact of disinformation on the public.

Therefore, we shall consider the extent to which the processes of reflection, refraction, polarization, diffraction, and interference can be applied to the visualization of the propagation of disinformation. In the cases of reflection and refraction, the following figure shows that as a wave encounters a barrier between two media, the part that is reflected does not change in wavelength although it changes in Amplitude and intensity. The refracted wave does change in wavelength although not in intensity. Since the refracted wave has a shorter wavelength, it will then have a lower speed value. In other words, the refracted wave slows down in medium 2.

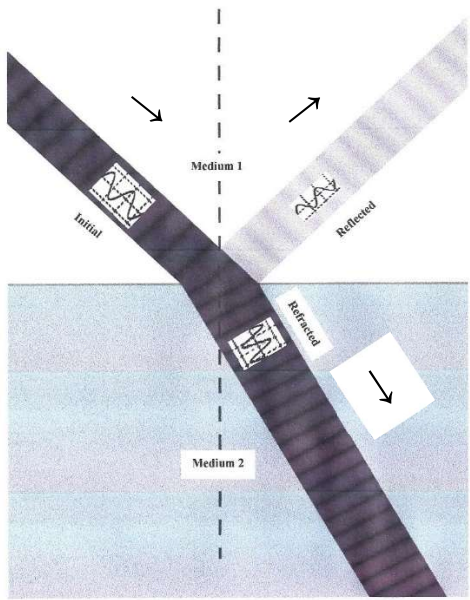

Figure 2. Reflection and refraction. An initially incident wave approaches through Medium 1 from the left, it is partly reflected into medium 1, and partly refracted into Medium 2. The reflected wave has a lower Amplitude (shown as lighter in shading) than the initial one, but the same wavelength. The refracted wave has the same Amplitude (equal shading), but shorter wavelength.

The compression of the space between the lines (representing the wavelength), and the change in direction of the arrows indicate that that the speed has decreased in Medium 2.

The process of diffraction is a result of a bending that waves experience when traveling through a medium as they encounter obstacles. This feature is more pronounced when the size of the objects that are the obstacles is comparable to the wavelength of the waves. Figure 3 shows this in addition to the fact that as waves go around obstacles their wavelengths do not change and neither do their speeds.

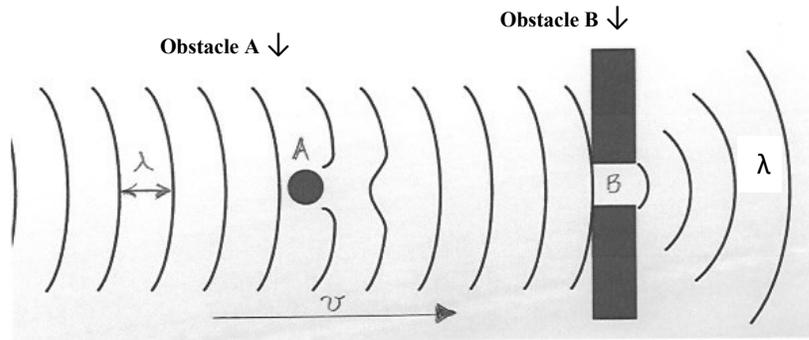

Figure 3. Diffraction of a wave traveling from the left and encountering objects A and B; as the wave bends around the obstacles, its wavelength and speed remain constant.

The interference of waves occurs when they combine and superimpose on one another, resulting in adding and subtracting or constructive and destructive interference. Figure 4 shows that the two waves combine and since they have the same Amplitude but differ in what is called their phase, they successively add up and cancel each other. The cancellation occurs when the two waves are totally opposite each other, or 180 degrees out of phase, the points along the equilibrium line in the figure.

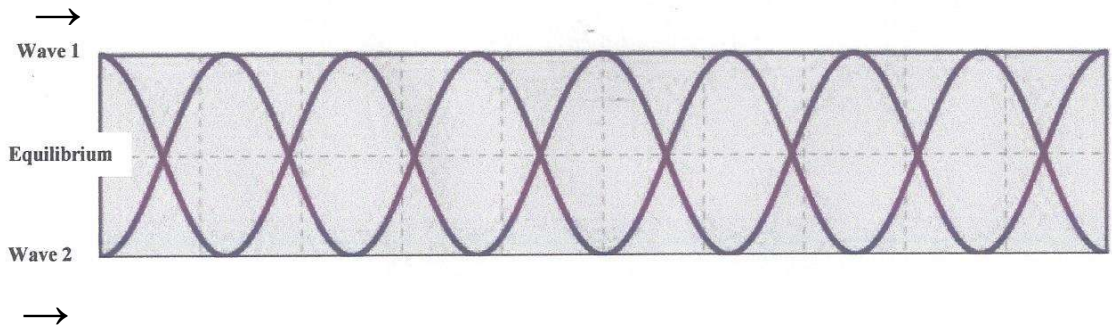

Figure 4. Two waves travel to the right and interact, as they successively add up and subtract, they are said to interfere constructively and destructively. The points along the Equilibrium line are the result of the two waves cancelling each other out.

Finally, the polarization of waves results from a filtering of some oscillations along specific directions. Figure 5 shows a wave from the left polarized along the vertical and the horizontal directions; passing through the vertical filter eliminates the horizontal component resulting in only the vertical one being transmitted.

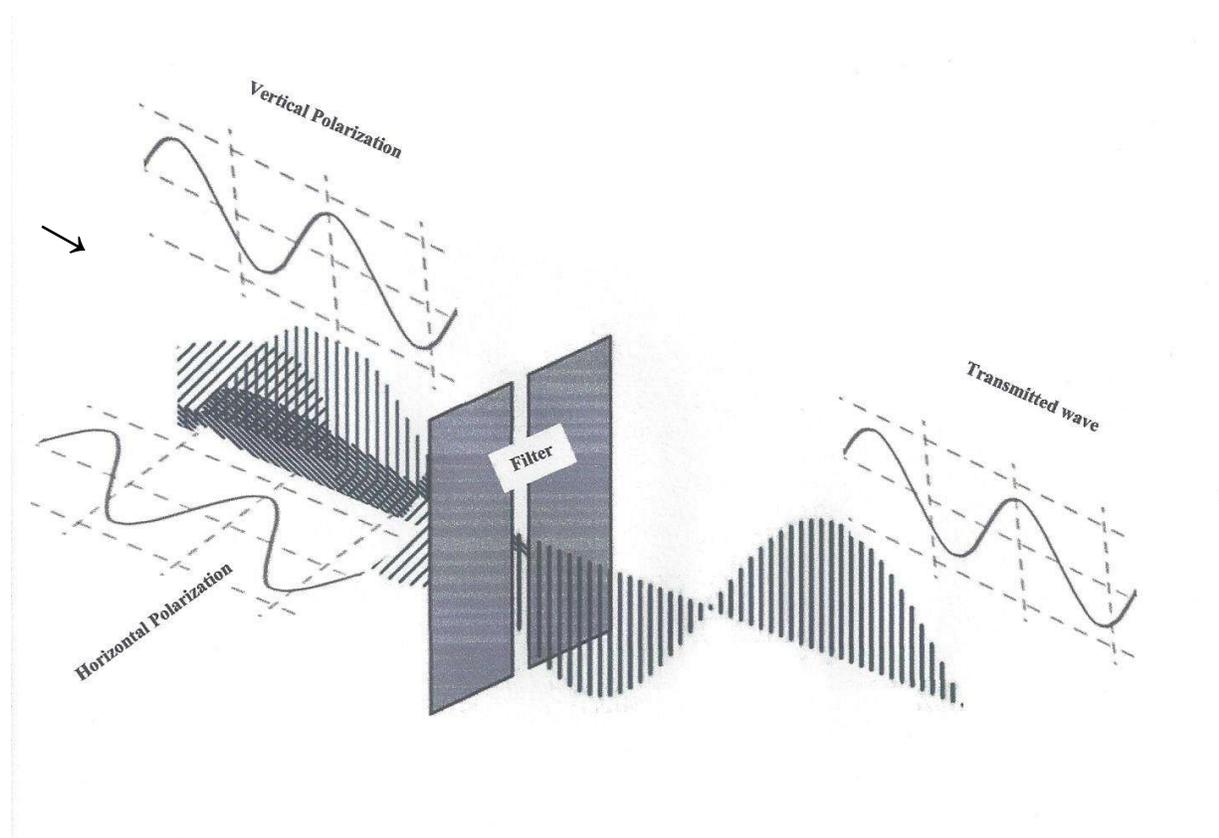

Figure 5. A wave approaching from the left exhibits horizontal and vertical polarization; after it passes through the filter, the horizontal component is blocked resulting in only the vertical one being transmitted.

The visualization of the propagation, behavior, and interactivity of disinformation as a wave entails the occurrence of all these processes. Our aim is not just to describe disinformation using wave properties, but to propose a way to deal with it using techniques that effectively neutralize and greatly reduce it. Consequently, the two fundamental objectives are to *slow down* the transmission of dissemination, and to *minimize* its impact.

## Mechanism

We define a medium as a setting or context for the dissemination of disinformation; therefore, the wave properties displayed by it as it travels through such a medium, or when it encounters another medium will replicate the general behavior of waves. We take the initial/natural medium (Medium 1) as that where disinformation propagates unimpeded, and its reception is one of implicit acceptance. In this regard, we can conclude that reflection, diffraction, and interference, which all occur in a medium, will not be useful for our purposes.

In reflection, the wave returns to the medium without its speed changing although its intensity can be somewhat diminished. This is analogous to what happens when disinformation silos or bubbles develop; if disinformation simply bounces back, believers just reinforce each other in their convictions. In diffraction, the bending of the waves does not change their wavelength and thus the speed remains the same, even if there is a deflection. Since disinformation behaves like a wave, in both cases we are unable to slow down its propagation. In the case of interference, the minimum

Amplitude and therefore the intensity of the resulting wave from the two that interact only occurs momentarily, or only at the points on the equilibrium line. An additional challenge when using interference to neutralize wave 1 taken to be the disinformation wave in Figure 4, requires that wave 2 completely cancels it. If we take wave 2 to represent a wave of true or accurate information, we must overcome the well-documented confirmation bias (Nickerson, 1998). This is the human predisposition to accept evidence that is supportive of our beliefs, and to discard that which contradicts them. It would be unrealistic to attempt to persuade individuals already convinced of the veracity of disinformation by providing them with an alternative and totally opposite point of view, exactly what is required to obtain total destructive interference as wave 2 must provide in interacting with wave 1 as they travel through the same medium.

We then concentrate on *refraction* and *polarization* as the means to develop a strategy for slowing down and significantly reducing the intensity or impact of disinformation. It must be acknowledged that dispersion and inverse-square dependence on the distance from the source of the disinformation waves also diminishes their intensity. However, these are purely natural consequences of its propagation, and we need interventions.

Hence the proposed two steps to effectively deal with disinformation as it displays the behavior of wave phenomena:

1. The speed of propagation can be decreased by providing a setting or context (Medium 2) that is *refractive*. This means that as shown in Figure 2, the transmitted wave changes direction and its wavelength is reduced (its reach) resulting in a lower speed. As in the physics definition, the refractive property of this medium is one where there is more resistance to the transmission of information than that found in the original (Medium 1) or initial context or setting.
2. The intensity of the signal or wave can be decreased by using a device (as a *polaroid filter*) that only allows for the transmission of selected oscillations or directions of wave polarization.

What do we mean by a refractive medium in a context other than the physical one? Since we are associating refraction with resistance, such a medium would differ from that originally chosen as the reference (e.g., air in optics) in the same transmissive way as all other materials differ from air. The reference criterion is that of a setting or context where the disinformation is unopposed or unexamined. Whenever the wave representing the disinformation penetrates this medium, where there is resistance (in the form of skepticism), its behavior will be different and its speed of propagation lower. What can be used as a filter in the processing of disinformation? The requirements of the provision of evidence, and its justification. The use of such a filter is empowering at the individual level, and it removes the stigma of censorship when such monitoring is left at the discretion of institutions and organizations.

The strategy provided by implementing the two steps that constitute a heuristic for an assessment of disinformation commonly referred to as 'fake news' can be implemented at any level and by anyone. All claims and assertions contained in the various types of information we come across must pass both tests before giving them any credence. They must be met with the same degree of skepticism and required level of evidence before acceptance. That will provide the necessary means to slow down the spread of disinformation by not blindly accepting it, and will also reduce its inherent persuasiveness by subjecting it to analysis.

There are attempts to address the issue of how to process information in educational settings, such as those at The Center for Media Literacy in California designed for K-12 students to become well informed news consumers. Nevertheless, similar approaches must be implemented in the larger public sphere so that individuals can effectively participate in a democracy.

## Conclusion

The ability to represent disinformation as wave phenomena based on its demonstrated analogous properties and behavior, can facilitate its management during this critical time of uncertainty about the veracity and accuracy of information being disseminated often in an indiscriminate manner by social media outlets. The ability to visualize how refraction and polarization can neutralize disinformation when understood as a wave, provides a means to slow down its propagation so that we may effectively control its impact on the public.

The epistemic reformulation of disinformation as a distinct phenomenon existing independently of our internal representations avoids the issues pertaining to freedom of expression and their related constitutionality. The benefits of such reification include opportunities for engaging in discussions within a setting that is not anthropocentric and thus less prone to the effects of ad hominem and affective features. This is particularly important in considerations of the role of conspiracies and the strong emotive component of conspiracy theories. As indicated, the use of polarization to describe a particular property of the disinformation being propagated, effectively removes its negative connotations when it is applied to the positions of individuals about issues surrounding such disinformation. The shift in the attribution of polarization, from individuals to external phenomena also helps in avoiding documented views of judgment of individual behavior and opinion that are based on dispositional rather than situational considerations (Gilbert & Malone, 1995).

In summary then, the use of a refractive-like medium and a polaroid-like filter provide the means to combat the proclivity to succumb to the confirmation bias, and the attribution theory. These human traits undermine our ability to think and act rationally by clouding our judgment.

The current impression among experts and many in the public about the impact of disinformation on the behavior of radicalized groups is an alarming one. If it is easy to understand the description of the impact of fake news on society as a public health crisis, it can also border on an epidemic. Therefore, if disinformation is easily described as something organic, why can't it be more effectively dealt with by interpreting its impact as mechanistic and equally counteracting it by means of a strategy that uses its own features as a defense mechanism?